\documentclass[12pt]{article}

\usepackage{amsfonts}
\usepackage{bbm}
\usepackage{bm}

\newcommand{\bmat}{\left(\begin{array}}
\newcommand{\emat}{\end{array}\right)}
\newcommand{\uno}{\mathbbm{1}}

\def\a{\alpha}

\def\b{\beta}

\def\d{\delta}

\def\-{\hphantom{-}}
\def\ov{\overline}
\def\s2{\frac{1}{\sqrt2}}

\def\oh{\frac{1}{2}}
\def\beq{\begin{equation}}
\def\eeq{\end{equation}}
\def\beqa{\begin{eqnarray}}
\def\eeqa{\end{eqnarray}}
\def\im{{\rm Im \,}}
\def\re{{\rm Re \,}}

\def\D{{\rm D}}
\def\T{{\rm T}}
\def\M{{\rm M}}
\def\K{{\rm K}}
\def\O{{\rm O}}
\def\S{{\rm S}}
\def\ads{{\rm AdS \,}}
\def\Z{{\mathbb Z}}
\def\C{{\mathbb C}}
\def\P{{\mathbb P}}

\def\ca{{\cal A}}
\def\cc{{\cal C}}

\def\cn{{\cal N}}

\def\cv{{\cal V}}

\def\cw{{\cal W}}

\def\cs{{\cal S}}

\def\deq#1{\mbox{$D$=#1}}
\def\neq#1{\mbox{$\cn$=#1}}
\def\Dsl{\,\raise.15ex\hbox{/}\mkern-13.5mu D} %this one can be subscripted

\newcommand{\msm}[1]{\mbox{\small$#1$}}
\newcommand{\inpar}[1]{\left(#1\right)}
\begin{document}

%----------------------------------------------------------------------%
%  numbering equations with section number
%----------------------------------------------------------------------%
\makeatletter
\@addtoreset{equation}{section}
\makeatother
\renewcommand{\theequation}{\thesection.\arabic{equation}}

%----------------------------------------------------------------------%
%  title page
%----------------------------------------------------------------------%

\pagestyle{empty}
%\vspace*{1.0in}
\rightline{IFT-UAM/CSIC-07-58}
%\rightline{\tt hep-th/0408036}
\vspace{10mm}
\begin{center}
\LARGE{A second look at \neq1  supersymmetric ${\rm AdS}_4$ vacua of type IIA supergravity 
\\[10mm]}
\large{ Gerardo Aldazabal${}^a$ and
Anamar\'{\i}a Font ${}^b$\footnote{Permanent address: Departamento de F\'{\i}sica, Facultad de Ciencias,
Universidad Central de Venezuela, A.P. 20513, Caracas 1020-A, Venezuela.} \\[5mm]}
\small{
${}^a$Instituto Balseiro - Centro At\'omico Bariloche, \\[-0.3em]
8400 S.C. de Bariloche, CNEA and CONICET, Argentina. \\[2mm] 
${}^b$Instituto de F\'{\i}sica Te\'orica UAM/CSIC,\\[-0.3em]
Facultad de Ciencias C-XVI, Universidad Aut\'onoma de Madrid, \\[-0.3em]  
Cantoblanco, 28049 Madrid, Spain 
\\[5mm]} 
\normalsize{\bf Abstract} \\[2mm]
\end{center}
{\normalsize
We show that a class of type IIA vacua recently found within the \deq4 effective approach 
corresponds to compactification on $\ads_4 \times \S^3 \times \S^3/\Z_2^3$.
The results obtained using the effective method completely match the general 
ten-dimensional analysis for the existence  of \neq1 warped compactifications 
on $\ads_4 \times \M_6$. 
In particular, we verify that the internal metric is nearly-K\"ahler and
that for specific values of the parameters the Bianchi identity of the RR 2-form is
fulfilled without sources. For another range of parameters, including the massless case,
the Bianchi identity is satisfied when D6-branes are introduced.
Solving the tadpole cancellation 
conditions in \deq4 we are able to find examples of appropriate sets of branes.
In the second part of this paper we describe how an example with internal space 
$\C\P^3$ but with non nearly-K\"ahler metric fits into the general analysis of flux vacua.
}

\newpage

%----------------------------------------------------------------------%
%  Resetting of counters
%----------------------------------------------------------------------%
\setcounter{page}{1}
\pagestyle{plain}
\renewcommand{\thefootnote}{\arabic{footnote}}
\setcounter{footnote}{0}
%----------------------------------------------------------------------%
%  Paper begins
%----------------------------------------------------------------------%

\section{Introduction}
\label{intro}

Four-dimensional \neq1 supersymmetric vacua of type II supergravity with fluxes can be
analyzed directly in \deq10 or by means of an effective potential formalism in \deq4. In this work
we point out that a class of type IIA vacua, 
with geometric fluxes switched on, that were found using the latter method \cite{cfi}
corresponds to compactification on $\ads_4 \times \S^3 \times \S^3/\Z_2^3$.
The results obtained using the effective formalism are in complete accord with the general conditions
for the existence  of $\ads_4 \times \M_6$ vacua \cite{lt, gmpt1, gmpt2}.
This is a particular example of the equivalence between the higher and lower dimensional approaches
considered lately in greater generality \cite{km1, Cassani}.

In the $\ads_4 \times\S^3 \times \S^3/\Z_2^3$ compactification, that we study in depth, 
we show that the internal metric is nearly-K\"ahler.  In \cite{bc} it was first
proven that when $\M_6$ is nearly-K\"ahler there are consistent vacua of massive
IIA supergravity with \neq1 supersymmetry in $\ads_4$. As also remarked in \cite{bc}, 
besides $\S^3 \times \S^3$,
there are other six-dimensional compact spaces that admit a nearly-K\"ahler metric, 
namely $\S^6$, $\C\P^3$ and $SU(3)/U(1)^2$ \cite{ssbook}.
However, these spaces are not group manifolds
and cannot be treated in a simple effective approach based on adding geometric
fluxes to a toroidal compactification. 
It would be interesting to formulate all nearly-K\"ahler compactifications
within the effective four-dimensional approach. A first step in this
direction is the Kaluza-Klein reduction on nearly-K\"ahler spaces \cite{Kashani}. 
The case of $SU(3)/U(1)^2$ has been considered in \cite{hp}.

A property of nearly-K\"ahler compactifications is that for special values
of the fluxes the Bianchi identity for the RR 2-form can be satisfied without 
adding sources \cite{bc, lt}. For other ranges of parameters it is necessary to add O6-planes,
D6-branes, or both, wrapping 3-cycles in the internal space. In any case,
including D6-branes is required to generate charged chiral multiplets.
In the $\S^3 \times \S^3/\Z_2^3$ compactification  we will present examples of supersymmetric 
D6-branes that can be included to fulfill the Bianchi identity or equivalently to cancel tadpoles.
This problem was first addressed in \cite{adhl} where it was argued that a certain setup
of D6-branes could cancel the tadpoles. We find similar results at the time we go further 
in proving tadpole cancellation because we supply the explicit background fluxes.

The second part of this paper is devoted to describing how other \neq1, 
$\ads_4$ vacua of massless IIA supergravity, discovered 
long time ago \cite{np, vst, stv, pp}, fit into the modern analysis of flux vacua. 
In these compactifications the internal space can be $\C\P^3$ or $SU(3)/U(1)^2$, 
but the metric is not nearly-K\"ahler. We will focus on the $\C\P^3$ example,
but the analysis can be easily extended to $SU(3)/U(1)^2$. We give explicit
expressions for the metric and the fluxes and then find the Killing spinor that
allows to derive the fundamental forms that define the $SU(3)$ structure. 

The organization of this paper is as follows. In section 2 we summarize the conditions for the
existence of \neq1 $\ads_4$ vacua derived from the \deq10 theory. We also discuss the issue 
of solving the Bianchi identity for the RR 2-form with or without sources. In section 3 
we study compactification on $\ads_4 \times \S^3 \times \S^3/\Z_2^3$ by describing the internal space 
in terms of a set of structure constants, the so-called geometric fluxes, known to give 
\neq1 vacua from the analysis of the \deq4 effective potential. We then explain how 
the Bianchi identity for $F_2$ can be satisfied in general by adding sources and present
as well a concrete configuration of D6-branes in the massless case.  
There is an important interplay with the results in the \deq4 effective formalism that
are collected in appendix A.         
Section 4 deals with the compactification on $\ads_4 \times \C\P^3$ that provides an example
where the internal space is not nearly-K\"ahler. In appendix B we show that the proposed
metric and background fluxes in $\C\P^3$ do satisfy the equations of motion and preserve \neq1 supersymmetry
in \deq4.

\section{Review of supersymmetric conditions in \deq10}
\label{d10}

We are interested in \neq1 compactifications of type IIA supergravity with fluxes turned on and
warped product geometry
\beq
ds^2 = e^{2A(y)} ds_4^2 + ds_6^2 \ ,
\label{geo}
\eeq
where $ds_4^2$ and $ds_6^2$ are respectively the line elements of $\ads_4$ and the 
internal compact space.
The general conditions that these vacua must fulfill were derived in \cite{lt} using Romans massive
action \cite{romans} and also in \cite{gmpt1, gmpt2} starting with the democratic formulation of 
IIA supergravity \cite{demo}. In this note we use the results and notation of \cite{gmpt2} that are 
more suited to compare with the effective potential approach.

By assumption, the internal manifold has strictly $SU(3)$ structure, i.e. it
admits only one nowhere vanishing invariant spinor which in turn allows to write a 
fundamental 2-form $J$ and a holomorphic 3-form $\Omega$ satisfying the relations
\beq
\Omega \wedge J=0 \quad ; \quad \Omega \wedge \Omega^* = -\frac{4i}3 J \wedge J \wedge J  \ .
\label{su3}
\eeq
In the most general supersymmetric solution of the equations of motion, the warp factor and the 
dilaton are constants related by $\phi=3A$. Moreover, the characteristic forms $J$ and $\Omega$ 
must meet the conditions   
\beq
dJ= 2\tilde m e^{-A} \re \widehat \Omega  \quad ; \quad
d\widehat \Omega =-\frac{4i}3 \tilde m e^{-A} J^2 -i\cw_2 \wedge J  
\ ,
\label{derj}
\eeq
where $\cw_2$ is a real primitive 2-form. Here 
$\widehat \Omega = -i e^{i(\a+\b)} \Omega$, with $\a, \b$, phases that enter in the 
normalization of the \deq10 supersymmetry parameters (see \cite{gmpt2} for more details).
The equations of motion also require $(\a - \beta)$ to be a constant.

Besides the constant $\tilde m$, the solutions depend on the IIA mass parameter $m$. These two
real quantities are combined into the complex constant  
\beq
\mu = e^{-i(\a-\b)}\, (m + i\tilde m) \ .
\label{mudef}
\eeq
The parameter $\mu$ enters in the covariant derivative of the \deq4 gravitino and it turns out
to be related to the cosmological constant through $\Lambda = -3 |\mu|^2$. This $\Lambda$  
is defined with respect to the unwarped ${\rm AdS}_4$ metric.

In the solution the field strengths are determined to be\footnote{The sign differences with respect to
equation (7.9) in \cite{gmpt2} are due to our conventions ${}^* J = J^2/2$ and ${}^*1 = J^3/6$,
where ${}^*$ is the Hodge dual in six dimensions.}
\beq
\begin{array}{lclcl}
H  = 2 m  e^{-A} \re \widehat \Omega  & \ ; \ & F_0 = - 5 m e^{-4A} 
& \ ; \ &
F_2 =  -e^{-3A} *d \im \widehat \Omega - 3\tilde m e^{-4A} J  \\
F_4  = -\frac32 m e^{-4A} J^2 &  \ ; \ &
F_6 = \frac12  \tilde m e^{-4A} J^3 &  .
\end{array}
\label{fluxsol}
\eeq
The relation to the NSNS and RR forms is given by
\beq
H = dB + \ov{H} \quad ; \quad 
F_p = d C_{p-1} - H \wedge C_{p-3} + \big(\ov{F} \wedge e^B\big) \left|_{p} \right.  \ .
\label{hfdef}
\eeq
The barred  quantities are background fluxes and $\ov{F} = \ov{F_0} +  \ov{F_2} + \ov{F_4} + \ov{F_6}$
is a formal sum.
  
Clearly, (\ref{derj}) implies $J \wedge dJ=0$ and $d(\re \widehat{\Omega})=0$.
This means that the internal space is always a half-flat manifold. 
If the torsion class $\cw_2$ vanishes
the internal space is nearly-K\"ahler and the RR 2-form simplifies to
\beq
F_2 = - \frac{\tilde m}3 \, e^{-4A} \, J  \ .
\label{f2nk}
\eeq 
This implies in particular that $dF_2 \not=0$ in nearly-K\"ahler compactifications. 

The Bianchi identities for $H$ and $F_4$ are automatically satisfied. On the other hand, for the
RR 2-form the generic results imply $dF_2-F_0H \not=0$. The situation is not hopeless because
there might be further contributions due to D6-branes or O6-planes wrapping 3-cycles in the internal
space. Actually, the Bianchi identity (BI) for $F_2$ is equivalent to tadpole cancellation 
conditions for the RR $C_7$ form that couples to such sources.

Following the prescription of \cite{gmpt2} we assume that the sources are smeared instead of localized.
This means that in the BI D6-branes and O6-planes can be represented by additional 3-forms in the internal 
space. This is actually the only consistent possibility for the $\ads_4$ vacua in which the
warp factor must be constant. Upon including smeared sources the BI becomes 
\beq
dF_2 - F_0 H + A_3 = 0 \ ,
\label{tadf2}
\eeq 
where $A_3$ is the Poincar\'e dual to internal 3-cycles wrapped by  D6-branes or O6-planes.
By virtue of (\ref{hfdef}), this identity can
be written purely in terms of background fluxes as $d\ov{F}_2 - F_0 \ov{H} + A_3 = 0$.

A property of the \neq1 $\ads_4$ vacua is that $H \propto dJ$. Thus, the form $A_3$ is necessarily
exact. In consequence, to saturate the Bianchi identity of the RR 2-form, or equivalently
to cancel $C_7$ tadpoles, the sources need not wrap non-trivial 3-cycles.
This point has been known for some time \cite{adhl, Cascales} and further elaborated
recently \cite{km2}. Due to the special properties of $\ads_4$ such D6-branes can still be stable.
  
When $F_0 \not=0$ there could be a solution of (\ref{tadf2}) without sources even if
$dF_2 \not=0$. Indeed, when the internal space is nearly-K\"ahler from the above results 
it follows that
\beq
dF_2 - F_0 H = \msm{\frac23} e^{-5A}(15m^2 - \tilde m^2)  \, \re \widehat \Omega \ .
\label{tadf3}
\eeq 
Therefore, it is possible to avoid sources, i.e. $A_3=0$, provided that $\tilde m^2 = 15 m^2$. 
This interesting fact was first obtained in \cite{bc} and later in \cite{lt}.
On the other hand, if $\tilde m^2 \not= 15 m^2$, sources must be added to fulfill
the Bianchi identity. For instance, if $\tilde m^2 > 15 m^2$ a solution can be achieved by adding
only D6-branes. This follows because supersymmetric 3-cycles are calibrated by $\re \Omega$
and in this case $\int_{\M_6}\re \Omega \wedge A_3 > 0$. Here we are taking 
$\widehat  \Omega = -i \Omega$ according to results in appendix A. 

It is also feasible to satisfy the Bianchi identity without sources and $F_0=0$ simply when 
$dF_2 =0$. Clearly, in this situation the internal space cannot be nearly-K\"ahler. Instead, the
torsion class $\cw_2$ must be non-zero. Examples of this type were actually found several
years ago \cite{np, vst, stv,pp}. In section \ref{ccp3} we discuss in detail the case of 
compactification on $\C\P^3$.

\section{Flux compactification on $\bm{{\rm AdS}_4 \times \S^3 \times \S^3}$}
\label{cs3s3}

We are interested in \neq1 type IIA vacua in presence 
of geometric fluxes $\omega^P_{MN}$ together with NSNS and RR fluxes. Such solutions 
can be viewed as compactifications in which the internal space 
has a basis of globally defined 1-forms satisfying
\beq
d\eta^P = -\oh \omega^P_{MN} \eta^M \wedge \eta^N \ ,
\label{gflux}
\eeq  
where the $\omega^P_{MN}$ are the structure constants of some Lie group $G$. 
If the Killing form $\K_{MN}=\omega^P_{MR} \omega^R_{NP}$ is non-degenerate,
$G$ is semisimple and furthermore it is compact if $\K_{MN}$ is negative definite.
If $G$ is not semisimple, but it has a discrete compact sub-group $\Gamma$, the
internal space can be compactified by taking the quotient $G/\Gamma$. This is
the case of the nil and solvmanifolds studied in \cite{gmpt2}. In this note
we rather study the situation where $G$ is compact and the internal space is
the $G$ group manifold. In particular, we want to show that in a class of supersymmetric
$\ads_4 \times \M_6$ vacua found in \cite{cfi} the structure constants are actually those of
$SU(2) \times SU(2)$ and the internal space is $\S^3 \times \S^3$ realized as
$SU(2) \times SU(2)\times SU(2)/ SU(2)_{\rm diag}$.
 
The number of independent geometric fluxes $\omega^P_{MN}$ can be reduced by imposing additional 
conditions on the internal space. We will enforce a $\Z_2\times \Z_2$ symmetry 
whose generators act as
\beqa
\Z_2 & : & (\eta^1, \eta^2, \eta^3, \eta^4, \eta^5, \eta^6) \to
(-\eta^1, -\eta^2, \eta^3, -\eta^4, -\eta^5, \eta^6) 
\nonumber \\[2mm]
\Z_2 & : & (\eta^1, \eta^2, \eta^3, \eta^4, \eta^5, \eta^6) \to
(\eta^1,-\eta^2, -\eta^3, \eta^4, -\eta^5, -\eta^6) \ .
\label{z2z2}
\eeqa
Furthermore, keeping in mind the eventual
need for orientifold planes to cancel tadpoles, the geometric fluxes are required 
to be invariant under an orientifold 
involution $\sigma$ which is also a $\Z_2$ symmetry given by 
\beq
\sigma : \eta^i \to \eta^i \quad ; \quad \eta^{i+3} \to -\eta^{i+3} \quad , \quad   i=1,2,3  \ .
\label{oaction}
\eeq
In the end only twelve geometric fluxes survive and they are further 
constrained by the Bianchi identities following from (\ref{gflux}). 
In the $\ads_4$ solutions found in \cite{cfi} there are only four independent parameters
$a$ and $b_i$ which appear in the structure equations 
\beq
\begin{array}{lcl}
d\eta^1 = -a \eta^{56} - b_1 \eta^{23} & \quad ; \quad &  d\eta^4 = - b_2\eta^{53} - b_3 \eta^{26} \\
d\eta^2 = -a \eta^{64} - b_2 \eta^{31} & \quad ; \quad & d\eta^5 = - b_1\eta^{34} - b_3 \eta^{61} \\
d\eta^3 = -a \eta^{45} - b_3 \eta^{12} & \quad ; \quad & d\eta^6 = - b_1\eta^{42} - b_2 \eta^{15} .
\label{abis}
\end{array}
\eeq
The notation $\eta^{12}=\eta^1 \wedge \eta^2$, etc. is understood.

For future purposes we record the 2, 3 and 4-forms invariant under the  $\Z_2\times \Z_2$ symmetry. 
These are
\beq
\begin{array}{lclclcl}
\omega_1 = - \eta^{14} & \quad ; \quad & \a_0 =\eta^{123} & \quad ; \quad & \b_0 =\eta^{456} & 
\quad ; \quad & \tilde \omega_1 = \eta^{2536} \\
\omega_2 = - \eta^{25} & \quad ; \quad & \a_1 =\eta^{156} & \quad ; \quad & \b_1 =\eta^{423} &
\quad ; \quad & \tilde \omega_2 = \eta^{1436} \\
\omega_3 = - \eta^{36} & \quad ; \quad & \a_2 =\eta^{426} & \quad ; \quad & \b_2 =\eta^{153} &
\quad ; \quad & \tilde \omega_3 = \eta^{1425} \\
&  & \a_3 =\eta^{453} & \quad ; \quad & \b_3 =\eta^{126} & \quad . \quad  &  
\label{allforms} 
\end{array}
\eeq
Notice that $\a_I$ and $\tilde \omega_i$ are even whereas  $\b_I$ and $\omega_i$ are odd under the
orientifold involution.
The normalization is 
\beq
\int_{\M_6} \a_i \wedge \b_j = 
\int_{\M_6} \omega_i \wedge \tilde \omega_j = \cv_6 \, \d_{ij} \ ,
\label{ccdef}
\eeq
where $\cv_6$ is a constant to be computed later on.
%zzz agregue on

When the geometric fluxes $a$ and $b_i$ are zero, the internal space can be compactified
into a flat six-dimensional torus. Moreover, the $\Z_2\times \Z_2$ symmetry that is assumed implies
that this torus is a product of three $\T_i^2$. Each 2-torus has a basis of 1-forms $(\eta^i, \eta^{i+1})$,
a K\"ahler modulus (area) $t_i$ and a complex structure parameter $\tau_i$ that must be real for consistency
with the orientifold involution. With this picture in mind we take the metric on $\M_6$, 
with $a, b_i \not=0$, to
still be given by
\beq
ds_6^2= \sum_{i=1}^3 \frac{t_i}{\tau_i} (\eta^i)^2 + t_i \tau_i (\eta^{i+3})^2  \ .
\label{metricm6}
\eeq
By construction, $t_i > 0$ and $\tau_i > 0$. 
Clearly, $\sqrt{g_6}= t_1 t_2 t_3$. Integrating gives the volume ${\rm Vol}(\M_6)= \cv_6 \, t_1 t_2 t_3$, 
where $\cv_6$ is the normalization constant defined above.

The hermitian almost complex structure corresponding to the metric is
\beq
J= - t_1 \eta^{14} - t_2 \eta^{25} - t_3 \eta^{36} = t_i \omega_i  \ .
\label{jm6}
\eeq
The associated holomorphic (3,0) form can be written as
\beq
\Omega = \sqrt{\frac{t_1t_2t_3}{\tau_1\tau_2\tau_3}} 
\, (\eta^1 - i\tau_1 \eta^4) \wedge (\eta^2 - i\tau_2 \eta^5) \wedge (\eta^3 - i\tau_3 \eta^6) \ .
\label{om6}
\eeq
These $J$ and $\Omega$ satisfy (\ref{su3}) so that they provide an $SU(3)$ structure on
the internal space $\M_6$.   
Notice also that under the orientifold involution, $J \to - J$ and $\Omega \to \Omega^*$.

{}From (\ref{abis}) we find that  $dJ$ and $d\Omega$ are not zero but $J \wedge dJ$ 
and $d(\im \Omega)$ do vanish. Thus, the $\M_6$ defined by (\ref{abis}) is a half-flat manifold.
Additional properties must be fulfilled for $\M_6$ to serve as internal space in 
an \neq1 supersymmetric $\ads_4$ vacua of type IIA. Moreover, it is necessary to turn on particular
NSNS and RR background fluxes. Now, from the discussion in \cite{cfi} we know 
that a solution is obtained with a precise set of fluxes invariant 
under the $\Gamma=\Z_2^3$ group of symmetries (\ref{z2z2}) and (\ref{oaction}).
Furthermore, in this solution the variables $t_i$ and $\tau_i$ that enter in
the metric satisfy specific relations. In the following our strategy is to use these
results to continue analyzing the properties of the $\M_6$ at hand.

In the appendix we review the conditions of \cite{cfi} to obtain
$\ads_4 \times \M_6$ supersymmetric minima. The fluxes allow a configuration with $t_1=t_2=t_3=t$,
where $t$ is completely fixed. A crucial property is that the structure constants $a$ and $b_i$ 
must all have the same sign. Also, the second equation in (\ref{realsu}) together with the 
explicit form of the moduli, c.f. (\ref{defrsu}), gives the very useful relations
\beq
b_i \tau_j \tau_k = 3a \quad \Rightarrow \quad \tau_i^2 = \frac{3 a b_i}{b_j b_k} \ , \quad
i \not= j \not= k \ .
\label{crux}
\eeq   
We then find
\beq
dJ = \frac32 \, \im (\cw_1 \Omega) \quad ; \quad
d \Omega =  \cw_1 \, J \wedge J  \quad ; \quad
\cw_1 =  \frac{2a}{\sqrt{t \tau_1 \tau_2 \tau_3}}
\label{djdo}
\eeq
In general the exterior derivatives of $J$ and $\Omega$ can be expressed in terms of torsion classes
(see e.g. \cite{grana}). In our case, from (\ref{djdo}) we easily see that the only non-zero class is
$\cw_1$. This is precisely the condition for the internal space to be nearly-K\"ahler.

It is a simple exercise to compute the Killing form for the structure constants given in (\ref{abis}).
We find
\beq
\K = -4 \, {\rm diag}(b_2b_3, b_1b_3, b_1 b_2, a b_1, a b_2, ab_3) \ .
\label{killf}
\eeq
Now, recall that to obtain $\ads_4 \times \M_6$ supersymmetric minima 
the geometric fluxes $a$ and $b_i$ must all have the same sign. 
Therefore, $\K$ is non-degenerate and negative-definite. We might guess that the semisimple compact algebra
being six-dimensional is that of $SU(2)\times SU(2)$. Indeed, after performing the
change of basis
\beq
\begin{array}{lcl}
\xi^1 = \sqrt{b_2 b_3} \, \eta^1 + \sqrt{ab_1} \, \eta^4 & \quad ; \quad &  
\hat \xi^1 = \sqrt{b_2 b_3} \, \eta^1 - \sqrt{ab_1} \, \eta^4 \\
\xi^2 = \sqrt{b_1 b_3} \, \eta^2 + \sqrt{ab_2} \, \eta^5 & \quad ; \quad &  
\hat \xi^2 = \sqrt{b_1 b_3} \, \eta^2 - \sqrt{ab_2} \, \eta^5 \\
\xi^3 = \sqrt{b_1 b_2} \,\eta^3 + \sqrt{ab_3} \, \eta^6 & \quad ; \quad &  
\hat \xi^3 = \sqrt{b_1 b_2} \,\eta^3 - \sqrt{ab_3} \, \eta^6 \ ,
\label{xibasis}
\end{array}
\eeq
the structure equations become
\beq
d\xi^i = -\frac12 \epsilon_{ijk} \, \xi^i \wedge \xi^j \quad ; \quad
d \hat \xi^i = -\frac12 \epsilon_{ijk} \, \hat \xi^i \wedge \hat \xi^j  \ .
\label{newstr}
\eeq
This confirms that the underlying algebra is that of $SU(2)\times SU(2)$.

We can take the $\xi^i$ and $\hat \xi^i$ to be two sets of $SU(2)$ left invariant 1-forms.
Concretely, 
\beqa
\hat \xi^1 & = & \cos \hat \psi d\hat \theta + \sin\hat \psi  \, \sin\hat \theta  \,d\hat \phi 
\nonumber \\[2mm]
\hat \xi^2 &= & -\sin \hat \psi d\hat \theta + \cos\hat \psi  \,\sin\hat \theta \, d\hat \phi 
\label{xihsu2} \\[2mm] 
\hat \xi^3& = & d\hat \psi + \cos \hat \theta \, d\hat \phi \ ,
\nonumber
\eeqa 
and similarly for the $\xi^i$. The range of angles is $0 \leq \hat \theta \leq \pi$,
$0 \leq \hat \phi \leq 2\pi$ and $0 \leq \hat \psi \leq 4\pi$. 

Our claim that the internal space is $\S^3 \times \S^3$ is supported by the explicit
form of the metric in the new basis. Substituting (\ref{xibasis}) into (\ref{metricm6})
readily gives
\beq
ds_6^2 = \frac{t}{\sqrt{3ab_1 b_2 b_3}}\, \big[ (\xi^i)^2 + (\hat \xi^i)^2 - \xi^i \hat \xi^i \big]  \ .
\label{s3s3metric}
\eeq   
This is an Einstein metric that belongs to a family of homogeneous metrics on $\S^3 \times \S^3$
\cite{gpp}. The isometry group is $SU(2)^3$ \cite{amv, aw}. There are two $SU(2)$'s from the left 
actions that leave $\xi^i$ and $\hat \xi^i$ separately invariant, and a further $SU(2)$ from a 
simultaneous right action by the same element on $\xi^i$ and $\hat \xi^i$. 
>From the metric and the explicit realization of the $SU(2)$ 1-forms 
the volume of $\S^3 \times \S^3$ can be evaluated to be
\beq
{\rm Vol}(\S^3 \times \S^3) = \frac{(4\pi)^4 \, t^3} {(4ab_1b_2b_3)^{3/2}} \equiv \cv_6 \, t^3  \ ,
\label{vols3s3}
\eeq
where $\cv_6$ is precisely the normalization constant introduced in (\ref{ccdef}).

In the new basis the fundamental forms $J$ and $\Omega$ are given by 
\beqa
J & = &  \frac{t}{2\sqrt{ab_1 b_2 b_3}} \, (\xi^1 \wedge \hat \xi^1 +  \xi^2 \wedge \hat \xi^2 +  
\xi^3 \wedge \hat \xi^3) 
\label{joebasis}  \\
\Omega & = & -\frac{t^{3/2}}{(3ab_1 b_2 b_3)^{3/4}} \, (\xi^1 + e^{2i\pi/3} \hat \xi^1) \wedge
(\xi^2 + e^{2i\pi/3} \hat \xi^2) \wedge (\xi^3 + e^{2i\pi/3} \hat \xi^3) \ .
\nonumber
\eeqa
Similar expressions have appeared in the literature some time ago \cite{adhl}
and more recently \cite{km2}.

At this point we must remember that our actual model is constrained by some
specific symmetries. Indeed, the geometric fluxes (\ref{abis}), as well as the NSNS and RR
backgrounds (\ref{fluxbg}), have been chosen to be invariant under the group $\Gamma=\Z_2^3$
of transformations given by the geometric $\Z_2 \times \Z_2$ (\ref{z2z2}) and the orientifold 
involution  $\sigma$ (\ref{oaction}). The action of $\sigma$ amounts to 
exchange of the spheres, $\xi^i \leftrightarrow \hat \xi^i$, which is clearly a symmetry
of the metric. On the other hand, the geometric $\Z_2 \times \Z_2$ corresponds to
\beqa
\Z_2 & : & (\xi^1, \xi^2, \xi^3, \hat \xi^1, \hat \xi^2, \hat \xi^3) \to
(-\xi^1, -\xi^2, \xi^3, -\hat \xi^1, -\hat \xi^2, \hat \xi^3) 
\nonumber \\[2mm]
\Z_2 & : & 
(\xi^1, \xi^2, \xi^3, \hat \xi^1, \hat \xi^2, \hat \xi^3) \to
(\xi^1, -\xi^2, -\xi^3, \hat \xi^1, -\hat \xi^2, -\hat \xi^3) 
\label{z2z2xi}
\eeqa 
which also leaves the metric invariant. The effect of these latter symmetries is to restrict
the range of the angles that define the 1-forms, c.f. (\ref{xihsu2}). The first and second $\Z_2$'s
imply respectively  $\hat \theta \equiv - \hat \theta$, and $\hat \psi \equiv - \hat \psi$
simultaneously with $\hat \phi \equiv - \hat \phi$, and analogous for the unhatted angles.
In the end we truly have internal space $\S^3 \times \S^3/\Gamma$, with volume given by
$\cv_6 t^3/8$. We will write
\beq
{\rm Vol}(\S^3 \times \S^3/\Gamma) = \cc \, t^3 \ ,
\label{vols3s3G}
\eeq
where $\cc=\cv_6/8=4\pi^4/(ab_1b_2b_3)^{3/2}$.

The nearly-K\"ahler metric on $\S^3 \times \S^3$ is also invariant under 
the order three transformation
\beq
\beta \ : \ \xi^i \to - \hat \xi^i \quad ; \quad
 \hat\xi^i \to \xi^i - \hat \xi^i  \ .
\label{betadef}
\eeq
This $\beta$-symmetry proves useful when studying properties of 3-cycles on
$\S^3 \times \S^3$ \cite{adhl}.

%%%%
\subsection{ D6-branes on  $\bm{\S^3 \times \S^3}$ and Bianchi identity for $\bm{F_2}$}

When $dF_2 \not=0$, the Bianchi identity for the RR 2-form can still 
be fulfilled by adding appropriate sources.  
The task is to find the 3-form $A_3$ that satisfies (\ref{tadf2}) and is the Poincar\'e dual of  
the 3-cycles wrapped by the sources.

In general, $A_3$ is some combination of the 3-forms of the internal space so that
it is important to characterize these forms, specially knowing that $A_3$ must be exact.
For $\S^3 \times \S^3$ the third Betti number is equal to two and the third
cohomology is rather simple. The two representative closed 3-forms are easier to describe in the
$(\xi^i, \hat \xi^i)$ basis. In fact, they are basically the volume forms of each $\S^3$, namely
\beq
h= \frac{\xi^{123}}{(4ab_1b_2b_3)^{3/4}}
\quad ;  \quad
\hat h= -\frac{\hat \xi^{123}}{(4ab_1b_2b_3)^{3/4}} \ .
\label{hhhdef}
\eeq
The normalization has been chosen so that
\beq
h \wedge \hat h = \frac{J^3}{6t^3} = \eta^{123456} \ .
\label{hnorm}
\eeq
{}From the six remaining 3-forms that can be constructed there are three exact combinations
given by $d(\xi^i \wedge \hat \xi^i)$. The corresponding forms in terms of the $\eta^M$
basis are found using the map (\ref{xibasis}). In particular, it follows that
\beq
a \eta^{456} = b_1 \eta^{423} = b_2 \eta^{153} = b_3 \eta^{126} = 
\left(\frac{ab_1b_2b_3}{64}\right)^{1/4} \!\! (h + \hat h) \ ,
\label{exf}
\eeq
where each equality is modulo exact forms.

Let us now study the homology. Our discussion resembles that in \cite{aw} and \cite{adhl}.
In $\S^3 \times \S^3$ we can identify three special 3-cycles as explained below.
\begin{trivlist}
\item[1.] The locus $\hat \xi^i =0$. By definition this is the first 3-sphere $\S_1^3$. From
the metric (\ref{s3s3metric}),
\beq
ds_6^2 \big|_{\hat \xi^i=0} = ds_3^2(\S_1^3) = \frac{t}{\sqrt{3ab_1 b_2 b_3}}\, (\xi^i)^2 \ .
\label{s1metric}
\eeq
>From the $\Omega$ form we find that $\im \Omega \big|_{\hat \xi^i=0}=0$, and moreover 
\beq
\re \Omega \big|_{\hat \xi^i=0}= -\frac{t^{3/2}}{(3ab_1 b_2 b_3)^{3/4}} \, \xi^{123}=
-{\rm dvol}(\S_1^3) \ .  
\label{reos1}
\eeq
This shows that the charge of a brane wrapping $\S_1^3$ is $-1$, it would
be an anti D6-brane in our conventions. For a D6-brane the 3-sphere must be wrapped in reverse
orientation. We will define the corresponding 3-cycle to be $D_1 = (-\S_1^3)$.

\item[2.] The locus $\xi^i =0$. By definition this is the second sphere $\S_2^3$. We now find
that
\beq
\re \Omega \big|_{\xi^i=0}= -{\rm dvol}(\S_2^3) \ . 
\label{reos2vol}
\eeq
Thus, a brane wrapping $\S_2^3$ has charge $-1$ and it is
an anti D6-brane in our conventions. Since $\im \Omega \big|_{\xi^i=0}=0$, we surmise
that the supersymmetric D6-brane must wrap the 3-cycle $D_2 = (-\S_2^3)$.
 
\item[3.] The locus $\xi^i= \hat \xi^i$. By definition this is the diagonal 3-sphere
$\S_D^3$. It is easy to check that $\im \Omega \big|_{\xi^i=\hat \xi^i}=0$.
Besides, from the metric (\ref{s3s3metric}) and the $\Omega$ form we deduce
\beq
\re \Omega \big|_{\xi^i=\hat \xi^i}= {\rm dvol}(\S_D^3) \ . 
\label{reosdvol}
\eeq
Due to some extra factors now there is a plus sign in front so that
the charge of a brane wrapping the diagonal 3-sphere is a D6-brane with charge $+1$.
We will denote  $D_0 = \S_D^3$.

The three 3-cycles discussed above, $D_0$, $D_1$ and $D_2$,  
cannot be independent since the third Betti number of $\S^3 \times \S^3$ is two.
In fact there is a linear relation among these cycles that will become clear when we discuss
the corresponding dual 3-forms.
\end{trivlist}

In general, given a 3-form $X$ integrated over one of the 3-cycles $D_i$, 
the Poincar\'e dual form $Y_i$ to $D_i$ in $\M_6=\S^3 \times \S^3$ is   
such that 
\beq
\int _{D_i}  \!\! X=\int _{\M_6}  \!\! X \wedge Y_i \ .
\label{poincaredef}
\eeq
For example, for $D_1=(-\S_1^3)$ we find
\beq
Y_1 = -\frac{\hat h}{\sqrt{\cv_6}} \ ,
\label{duald1}
\eeq
where $\hat h$ is defined in (\ref{hhhdef}).
To demonstrate this we can choose
\beq
X= {\rm dvol}(D_1)=-\frac{t^{3/2}}{(3ab_1 b_2 b_3)^{3/4}} 
\, \xi^{123}=- \frac{V_3}{(4\pi)^2}\xi^{123} \ ,
\label{xf1}
\eeq
so that $\int _{D_1} X= V_3$. On the other hand we can also compute
\beq
\int _{\M_6} \!\! X \wedge \left(-\frac{\hat h}{\sqrt{\cv_6}}\right) = V_3 \ .
\label{checkduald1}
\eeq
In a similar fashion we obtain the dual to $D_2=(-\S_2^3)$ to be
\beq
Y_2 = -\frac{h}{\sqrt{\cv_6}} \ ,
\label{duald2}
\eeq
where $h$ is defined in (\ref{hhhdef}).

We can now compute the intersection number of the 3-cycles $D_1$ and $D_2$
by means of the representative dual 3-forms. This is
\beq
D_2\cdot D_1 =\int_{D_1} Y_2 = \int_{\M_6}Y_2 \wedge Y_1 = 
\frac{1}{\cv_6}  \int_{\M_6} h \wedge {\hat h}=1 \ .
\label{int12}
\eeq
This agrees with the analysis of \cite{aw}. 

We still need to find the dual 3-form of the diagonal 3-sphere $D_0$. In this case it is
convenient to use the $\eta^M$ basis. We notice that $\xi^i = \hat \xi^i$ amounts to
going to the locus $\eta^4=\eta^5=\eta^6=0$. Either by changing variables or by evaluating
directly in (\ref{om6}), we obtain
\beq
{\rm dvol}(D_0)= \frac{t^{3/2}}{\sqrt{\tau_1\tau_2\tau_3}} 
\, \eta^{123} \ . 
\label{dvoldeta}
\eeq
It then follows that the dual 3-form is given by
\beq
Y_0 =\frac{a(4ab_1b_2b_3)^{1/2}}{4 \pi^2}\,  \eta^{456} \ ,
\label{duald0}
\eeq
where we have used that $\tau_1\tau_2\tau_3=(27/ab_1b_2b_3)^{1/4}$ as implied by (\ref{crux}).

As mentioned before, there must be a linear relation among the three supersymmetric 3-cycles
that have been identified. The claim is that
\begin{equation}
 D_0+D_1+D_2= 0 \ ,
\label{linD}
\end{equation}
in homology.
This can be simply understood in terms of the dual 3-forms. In fact, from (\ref{exf})
we have $Y_0= \frac{h+\hat h}{\cv_6}$, up to exact forms. Therefore, in cohomology, 
$Y_0+Y_1+Y_2= 0$, 
modulo exact forms. This confirms the validity of (\ref{linD}).

The remaining intersection numbers are also easily calculated. We find for instance
$D_0\cdot D_2 = \int_{\M_6}Y_0 \wedge Y_2 =1$. In general, 
\beq
D_i \cdot D_j = \int_{\M_6}Y_i \wedge Y_j = \delta_{j,i-1} -\delta_{j,i+1} \ ,
\label{intij}
\eeq
where the indices are defined modulo 3. 
These are the intersection numbers found in \cite{aw}. 
In particular they satisfy, $D_i\cdot ( D_0+D_1+D_2)=0$, consistent with
(\ref{linD}).

We will now carry the discussion in the quotient space
$\S^3\times \S^3/\Gamma$ with $\Gamma=\Z_2^3$.
To the 3-cycles, $D_i$ in the covering space we associate $D_i^\prime$ with corresponding
dual forms $Y_i^\prime$ in the quotient.
Closely following \cite{aw}, let us assume that the lifting to the covering space 
$\M_6=\S^3\times \S^3$ is given by
the map  
\beqa
(Y_0^\prime,Y_1^\prime, Y_2^\prime) &\rightarrow & (Y_0,  8 Y_1, 8 Y_2) 
\nonumber \\[2mm]
(D_0^\prime, D_1^\prime,D_2^\prime) &\rightarrow & (D_0, 8 D_1, 8 D_2)  \ .
\label{lifti}
\eeqa
With this Ansatz we then obtain for instance,
\beqa
D_2^\prime \cdot D_1^\prime & = &
\int_ {\cs_6} \!\! Y_2^\prime \wedge Y_1^\prime 
 =  \int_ {\cs_6} \!\! 8Y_2 \wedge 8 Y_1 \ = \
\frac18 \int_ {\M_6} \!\!\! 8Y_2 \wedge 8Y_1 =8
\nonumber \\[2mm]
D_0^\prime \cdot D_2^\prime & = &
\int_ {\cs_6} \!\! Y_0^\prime \wedge Y_2^\prime  =  
\int_ {\cs_6} \!\! Y_0 \wedge 8Y_2 \ = \
\frac18 \int_ {\M_6} \!\! Y_0 \wedge 8Y_2 =1 \ ,
\label{intsq}
\eeqa
where we have defined $\cs_6=\M_6/\Gamma=\S^3\times \S^3/ \Gamma$ to streamline expressions.
As expected, this is consistent with the normalization
\beq
\int_ {\cs_6}\!\! \eta^ {123456}=\frac{\cv_6}{8}  \,={\cc}
\label{MGvolume}
\eeq
where ${\cc}t^3$ is the volume of $\cs_6$. 
We will see that these intersection numbers also arise in our model 1 in \deq4
discussed in appendix A.

According to \cite{aw}, the 3-cycle $D_0^\prime$ corresponds to $D_0^\prime= S_D^3/{\Gamma}$. 
Namely, $D_0=S_D^3$ is an 8-fold cover of $D_0^\prime$.
Since cycles are not independent, this indicates that wrapping $N$ D6-branes around each
of the cycles $D_i^\prime$ with $i=1,2$, requires wrapping  $D_0^\prime$ $8N$ times. In other words,
\beq
 8D_0^\prime+D_1^\prime+D_2^\prime= 0 \ ,
\label{linDp}
\eeq
which is true by virtue of the map (\ref{lifti}) and the relation (\ref{linD}).

With all the information collected so far we can already establish a
connection to our model 1 explained in appendix A. In this model, with mass parameter $F_0=0$, 
we found that tadpoles could be cancelled by a setup of supersymmetric D6-branes wrapping 
particular factorizable 3-cycles in the $\eta^M$ basis.
The concrete configuration is summarized in table \ref{adsm0} where the 3-cycles are
explicitly given. It consists of a stack of $8N_B$ D6-branes wrapping a cycle $\Pi_A$,
$N_B$ D6-branes wrapping a cycle $\Pi_B$, plus $N_B$ D6-branes wrapping the mirror
cycle $\widetilde{\Pi}_B$. In the model, the geometric flux parameters satisfy
$a=b_i=2N_B/c$, where $c$ is related to the RR 2-form background.
Interestingly enough,  it is possible to represent these factorizable cycles
in terms of the supersymmetric 3-cycles in $\S^3\times \S^3/\Gamma$.
In fact, the following identifications can be made
\beq
\Pi_A= (1,0)^3 \equiv D_0^\prime
\quad ; \quad
\Pi_B = (-1,1)^3 \equiv D_2^\prime
\quad ; \quad 
\widetilde{\Pi}_B=(-1,-1)^3 \equiv D_1 ^\prime
\label{idenmod1} 
\eeq
Evidence for these matchings comes from the equivalence of the loci described in both
the $\eta^M$ and the $(\xi^i,\hat \xi^i)$ basis, and from agreement of the intersection numbers. 
For instance, $\Pi_A \cdot \Pi_B =1 = \D_0^\prime \cdot \D_2^\prime$ and
$\Pi_B \cdot \widetilde{\Pi}_B = 8 = \D_2^\prime \cdot \D_1^\prime$.
Besides, below we will check that the corresponding dual 3-forms do coincide.

Based on the above results from the analysis of supersymmetric 3-cycles in 
$\S^3\times \S^3/\Gamma$ we conclude that
a setup of D6-branes wrapping the cycles $D_0^\prime$, $D_1^\prime$ and $D_2^\prime$,
will lead to tadpole cancellation. Otherwise stated, 
the corresponding dual 3-forms must add up to the precise 3-form $A_3$ needed to saturate the
Bianchi identity. To substantiate this claim 
we will examine the Bianchi identity for the RR 2-form in more detail.
The starting point is equation (\ref{c7tadpole}). 
For sources wrapping space-time
the RR 7-form can be written as $C_7={\rm dvol}_4 \wedge X$, where $X$ is some 3-form in
the internal space which we take to be $\cs_6=\S^3\times \S^3/\Gamma$.
Then, (\ref{c7tadpole})
leads to   
\beq
\int_{\cs_6}\hspace{-2mm} X \wedge (d\ov{F}_2 - F_0 \ov{H}) +
\sqrt{\cc}\, \sum_a N_a Q_a \int_{\Pi_a} \hspace{-2mm} X = 0 \ .
\label{c7tadpolevol}
\eeq
The factor of $\sqrt\cc$ is necessary because we are writing $d\ov{F}_2$ and $\ov{H}$ in a basis
of forms with normalization (\ref{MGvolume}) or analogous in terms of the $(\xi^i, \hat \xi^i)$
1-forms. 

To continue, recall that $\int_{\Pi_a} \hspace{-2mm} X = \int_{\cs_6} \hspace{-1mm} X \wedge Y_a^\prime$,
where the 3-form $Y_a^\prime$ is the Poincar\'e dual of the 3-cycle $\Pi_a$. 
Thus, from the above integral we arrive at the BI
\beq
d\ov{F}_2 - F_0 \ov{H} +
\sqrt{\cc}\, \sum_a N_a Q_a  Y_a^\prime = 0 \ .
\label{biovo}
\eeq
In terms of the notation in section \ref{d10} we have
\beq
A_3 =  \sum_a N_a Q_a \, A_3^a \ ,
\label{a3a}
\eeq 
where $A_3^a=\sqrt{\cc}\, Y_a^\prime$ is the contribution of each individual source. 
Recall that $N_a$ is the number of D6-branes or O6-planes wrapping the 3-cycle $\Pi_a$
and $Q_a$ is the corresponding charge. 

In the following we focus on the massless case $F_0=0$ as in model 1 of appendix A.
As argued in section \ref{d10}, when $m=0$,  necessarily
sources of positive charge must be included to satisfy the BI. In this case, 
in our $\S^3\times \S^3/\Z_2^3$
compactification, from previous results we know that $d\ov{F}_2$ is given by
\beq
d\ov{F}_2  =  - \frac{c}{t} dJ =  -\frac {3c}{2t} \cw_1 \im \Omega \ . 
\label{bgf2}
\eeq
In the $\eta^M$ basis this yields the rather simple expansion
\beq
d\ov{F}_2 
=  -c(3a\eta^{456}- b_1 \, \eta^{423}-b_2 \, \eta^{153}- b_3 \, \eta^{126})  \ .
\label{bianchimzero}
\eeq
Our results for tadpole cancellation in model 1 in appendix A suggest a solution to
the BI, $d\ov{F}_2 + A_3=0$. 
Concretely we propose that in this situation 
$A_3$ can be written as
\beq
A_3 = N_B(8 A_3^A + A_3^B + \widetilde A_3^B)  \ ,
\label{a3sum}
\eeq
because $N_A=8N_B$ and $Q_A=Q_B=1$. Indeed, it is straightforward to check that
the BI is satisfied with  
\beqa
A_3^A & = &  \eta^{456} \ , \nonumber \\
A_3^B & = &  -\big(\eta^{456} + \eta^{423} +
\eta^{153} +  \eta^{126} + \eta^{123} + \eta^{156}  +
\eta^{426} + \eta^{453} \big) \ ,
\label{a3m0} \\
\widetilde A_3^B & = &  -\big(\eta^{456} + \eta^{423} +
\eta^{153} +  \eta^{126} - \eta^{123} - \eta^{156}  -
\eta^{426} - \eta^{453} \big) \ ,
\nonumber
\eeqa
as long as $a=b_i=2N_B/c$, which precisely guarantees tadpole cancellation.

To close our argument we compare the dual 3-forms $Y_a^\prime$ with those found before
for the supersymmetric 3-cycles in  $\S^3\times \S^3/\Z_2^3$. We find
\beqa
Y_A^\prime & = &  \frac1{\sqrt\cc} A_3^A = \frac{(ab_1b_2b_3)^{3/4}}{2 \pi^2}\,  \eta^{456} = 
Y_0 =  Y_0^\prime
\nonumber \\[2mm]
Y_B^\prime & = &  \frac1{\sqrt\cc} A_3^B =  8 (-\frac{ h}{\sqrt\cv_6})= 8 Y_2= Y_2^\prime
\label{alleluja} \\
\widetilde{Y}_B^\prime & = &  \frac1{\sqrt\cc} \widetilde {A}_3^B =  8 ( -\frac{\hat h}{\sqrt\cv_6})=
8 Y_1= Y_1^\prime
\ . 
\nonumber
\eeqa
Therefore, the cycles wrapped by D6-branes correspond to the ``quotient spheres''
$D_0^\prime$, $D_1^\prime$ and $D_2^\prime$, as already anticipated in (\ref{idenmod1}).

As we might suspect, a more generic solution to the BI can be obtained as we now
explain. Again in the massless case, the problem is to solve
\beq
d\ov{F}_2 + 
\sqrt{\cc}\, \sum_a N_a Q_a  Y_a^\prime = 0 \ .
\label{biovo2}
\eeq
In general we can attempt a solution with 3 stacks of D6-branes wrapping 
the supersymmetric quotient 3-spheres so that
\beq
A_3 = \sqrt{\cc}\, \sum_a N_a Q_a  Y_a^\prime = 
\sqrt{\cc}\, (N_0 Y_0^\prime+ N_1 Y_1^\prime+ N_2 Y_2^\prime) \ ,
\label{a3generic}
\eeq
setting the charges to 1. Now, as suggested by (\ref{linDp}), we choose $N_0=8N$,
$N_1=N_2=N$. Then,
\beq
A_3  =  
8\sqrt{\cc}\, N( Y_0+  Y_1+  Y_2) 
 =  \frac {2 N}{(ab_1b_2b_3)^{\frac14}}
(3a\eta^{456}- b_1 \, \eta^{423}-b_2 \, \eta^{153}- b_3 \, \eta^{126}) \ ,
\label{a3generic2}
\eeq
where we used the lifting (\ref{lifti}) and the expansions of the dual forms
in the $\eta^M$ basis. Comparing with (\ref{bianchimzero}) we see that the BI is satisfied
provided that
\begin{equation}
 c= \frac {2 N}{(ab_1b_2b_3)^{\frac14}} \ .
\label{cgeneric}
\end{equation}
In the \deq4 formulation developed in section A.1, this generic solution can be associated to
a particular configuration of supersymmetric D6-branes similar to model 1.
The setup consists of $N_B$
D6-branes wrapping $\Pi_B=(-1,k) \otimes (-1,\ell) \otimes (-1,m)$, where
$(k,\ell,m)$ are positive integers, $N_B$ D6-branes along the mirror 3-cycle $\widetilde{\Pi}_B$,
plus $N_A=8N_B$ D6-branes wrapping $\Pi_A=(1,0)^3$.
It is not difficult to check that tadpoles are cancelled, and $\Pi_B$ is supersymmetric, as
long as $ac=2N_B$, $b_1c=2N_B \ell m$, $b_2c=2N_B k m$ and $b_3c=2N_B  k \ell$.
Combining these parameters we reproduce (\ref{cgeneric}) with $N=N_B \sqrt{k\ell m}$.

To finish this section we would like to comment on the massless spectrum originating from
the configuration of D6-branes. The interpretation is that
in $\S^3\times \S^3/\Gamma$
a setup of $N_B$ D6-branes wrapping each of the cycles $D_1^\prime$ and $D_2^\prime$, as well
as $N_A=8N_B$ D6-branes wrapping $D_0^\prime$, allows to satisfy the BI. 
These D6-branes produce an anomaly-free spectrum with
gauge group $U(N_A) \times U(N_B) \times U(N_B)$ and massless matter content 
\beq
({\bm{N_A}},{\bm{\ov{N_B}}},{\bf 1}) + ({\bm{\ov{N}_A}},{\bf 1},{\bm{N_B}}) 
+ 8({\bf 1},{\bm{N_B}},{\bm{\ov{N}_B}}) \ ,
\label{specinters}
\eeq 
consistent with the intersection numbers of the 3-cycles.
Notice that the spectrum is chiral and, therefore it cannot be continuously deformed away. 
This signals the stability of the D6-brane configuration.

The above spectrum is the same as in model 1 in appendix A.
We are assuming that the curvature of the 3-spheres wrapped by the D6-branes is large.
In fact, the radius is controlled by the size modulus $t$ whose vev can turn out large,
for instance by adjusting the RR flux $e_0$ \cite{cfi}. 
On the other hand, the fact that the D6-branes wrap 3-spheres can have
interesting consequences. For instance, since the first Betti number of $\S^3$
is zero, open string massless scalar moduli are not expected.
In the lines of \cite{cg} these, adjoint, scalars would become massive through 
$\mu $ terms in the effective superpotential\footnote{We thank 
P.~C\'amara for these observations.}.
This could be an appealing feature from a phenomenological perspective.

So far we have concentrated here in massless type IIA without orientifold planes. 
Extensions to more general cases can in principle be worked out
and could lead to attractive models from the phenomenological point of view.

\section{Flux compactification on $\bm{{\rm AdS}_4 \times \C\P^3}$}
\label{ccp3}

Compactification of massless type IIA supergravity on  ${\rm AdS}_4 \times \C\P^3$
have been studied in detail in  \cite{np, vst, stv}. The idea was to look for solutions similar
to the Freund-Rubin compactification of eleven-dimensional supergravity. Thus, a non-trivial
background for the RR 4-form, $F_4 \propto {\rm dvol}_4$ is turned on. By Hodge duality
this is equivalent to $F_6 \propto {\rm dvol}_6$. The solutions are unwarped and have
constant dilaton. There is no $H$ flux.
The RR 2-form flux can be chosen to be $F_2 \propto J$, where $J$ is the
fundamental form of $\C\P^3$. When the internal metric is given by the Fubini-Study metric
the equations of motion are satisfied. Furthermore the Bianchi identity for $F_2$ is automatic because
$dJ=0$. It can be shown that an extended \neq6 supersymmetry is preserved.

Applying the general results reviewed in section \ref{d10} we can see that for $m=0$ there is
a solution with \neq1 supersymmetry
when the metric in  $\C\P^3$ is chosen to be nearly-K\"ahler. However, in this case the Bianchi identity
for $F_2 \propto J$ is not satisfied because $dJ \not=0$. Presumably the tadpoles could be cancelled by
adding D6-branes. The third homology of $\C\P^3$ is trivial but there could exist supersymmetric
3-cycles.  

Yet another \neq1 solution with $m=0$ can be found by choosing the metric on $\C\P^3$ 
and the RR 2-form flux to descend from the metric of the
squashed seven-sphere which gives an \neq1 solution of \deq11 supergravity \cite{dnp}.
In this case the $\C\P^3$ metric is not Einstein and therefore not
nearly-K\"ahler. According to the general analysis it must be that the metric
is such that the two torsion classes $\cw_1$ and $\cw_2$ are different from zero. In fact
setting $\widehat\Omega=-i\Omega$ in (\ref{derj}) tells us that
\beq
dJ= \frac32 \cw_1 \im \Omega  \quad ; \quad d\Omega= \cw_1 J^2 + \cw_2 \wedge J ,
\label{nonnk}
\eeq  
where $\cw_1=\frac43 \tilde m e^{-A}$ and $\cw_2$ is a real primitive 2-form. 
Substituting in (\ref{fluxsol}) then gives
\beq
F_2= -\frac 14 \cw_1 J + {}^*(\cw_2 \wedge J) \ ,
\label{f2nnk}
\eeq
where we have put the warp factor to zero. In principle it is then feasible to attain
$dF_2=0$ even if $dJ\not=0$. Below we try to check these claims.  

The generic metric on  $\C\P^3$ can be constructed as a bundle with base $\S^4$ and
fiber $\S^2$. Denoting by $(\theta, \varphi)$ the coordinates of the $\S^2$ this means that   
\beq
ds_6^2 = d\tilde s_4^2
+  \lambda^2 (d\theta - \sin \varphi \ca^1 + \cos\varphi \ca^2)^2
+ \lambda^2 \sin^2 \theta (d\varphi - \frac{\cos\theta}{\sin\theta}(\cos \varphi \ca^1 
+ \sin\varphi \ca^2) + \ca^3)^2
\ , 
\label{cp3metric} \\
\eeq
where $d\tilde s_4^2$ is the line element of $\S^4$ and $\ca^A$ is the self-dual $SU(2)$
instanton potential on $\S^4$. More explicitly,
\beq
d\tilde s_4^2 = d\psi^2 + \frac14 \sin^2 \psi \, \Sigma^A \Sigma^A
\quad ; \quad
\ca^A = \cos^2 \frac{\psi}2 \, \Sigma^A \ .
\label{s4mi} 
\eeq
The $\Sigma^A$, $A=1,2,3$,  are left-invariant $SU(2)$ 1-forms 
for which we use coordinates
\beqa
\Sigma^1 & = & \cos \gamma \, d\a + \sin \gamma \sin \a \, d\b \ , 
\nonumber \\
\Sigma^2 & = & -\sin \gamma \, d\a + \cos \gamma \sin \a \, d\b \ , 
\label{s3forms} \\
\Sigma^3 & = & d\gamma + \cos \a \, d\b \ , 
\nonumber 
\eeqa
Notice that $d\Sigma^A = -\oh \epsilon_{ABC} \Sigma^B \wedge \Sigma^C$.

In the following we will employ a flat Sechsbein defined as
\beqa
e^1 & = & d\psi \quad ; \quad e^j = \oh \sin \psi \, \Sigma^{j-1} \ , \ j=2,3,4  \ , 
\nonumber \\[2mm]
e^5 & = & \lambda(d\theta - \sin \varphi \ca^1 + \cos\varphi \ca^2)  \ ,
\label{bein6} \\[2mm] 
e^6 & = & \lambda\sin\theta (d\varphi - \frac{\cos\theta}{\sin\theta}(\cos \varphi \ca^1 + 
\sin\varphi \ca^2) 
+ \ca^3) \ .
\nonumber
\eeqa
In the flat basis the Ricci tensor of the $\C\P^3$ metric is diagonal with components
\beq
R_{ab}=(3-\lambda^2) \, \d_{ab}  \quad ; \quad 
R_{ij}=(\lambda^2 + \frac1{\lambda^2}) \, \d_{ij} \ ,
\label{riccicp3}
\eeq
where $a,b=1, \cdots, 4$, and $i,j=5,6$.

Taking $\lambda^2=1$ gives the standard Einstein metric that is K\"ahler. 
A second Einstein metric that is nearly-K\"ahler is obtained setting $\lambda^2=\frac12$.
In both cases the Einstein equation of motion of type IIA supergravity can be solved
with $F_2 \propto J$.
Another solution can be found choosing $\lambda^2=\frac15$ and turning on an
appropriate RR 2-form flux. Concretely, $F_2$ must be
\beq
F_2 = -\lambda \sin \theta \sin \varphi (e^{12} + e^{34}) - 
\lambda \sin \theta \cos \varphi (e^{13} + e^{42}) -
\lambda \cos \theta (e^{14} + e^{23}) - 
\frac1{\lambda} e^{56}  \ .
\label{f2cp3}
\eeq
It can be checked that $dF_2=0$ and $\nabla_m F^{mn}=0$. Moreover, we will see that $F_2$ is of the 
expected form (\ref{f2nnk}), with $\cw_2 \not=0$. In appendix B we will check that all 
equations of motion are satisfied and that \neq1 supersymmetry is preserved.    

As already stressed in \cite{np, vst, stv}, the new $\C\P^3$ compactification of massless
IIA supergravity is directly related to compactification of \deq11 supergravity on the squashed
seven-sphere \cite{dnp}. Indeed, the metric on the squashed $\S^7$ can be written 
as\footnote{The metric on the squashed $\S^7$ is 
$ds_7^2 = d\tilde s_4^2 + \lambda^2( d\sigma^A - \ca^A)^2$, where $\sigma^A$ is a second set
of $SU(2)$ left-invariant 1-forms. To recover (\ref{ss7}) just set
$\sigma^1 =  \sin \varphi \, d\theta + \sin \theta \cos \varphi \, d\tau$, 
$\sigma^2 =  -\cos \varphi \, d\theta + \sin \theta \sin \varphi \, d\tau$, 
$\sigma^3 =  -d\varphi  + \cos \theta \, d\tau$.} 
\beq
ds_7^2 = (\lambda d\tau - A)^2 + ds_6^2 \ ,
\label{ss7}
\eeq
where $ds_6^2$ is the above metric on $\C\P^3$ and the gauge potential $A$ is such that
$dA$ gives precisely the RR 2-form background displayed in (\ref{f2cp3}). When  
$\lambda^2=\frac15$ this seven-dimensional metric is Einstein and admits only one Killing spinor. 

The fundamental forms $J$ and $\Omega$ can be obtained from the Killing spinor in six dimensions.
Details are presented in appendix B. The main results are
\beqa
J & = & 
-\sin \theta \sin \varphi (e^{12} + e^{34}) - \sin \theta \cos \varphi (e^{13} + e^{42}) -
\cos \theta (e^{14} + e^{23}) + e^{56}  \ ,
\nonumber \\[2mm]
\re \Omega & = &  \cos\theta \cos\varphi (e^{126} + e^{346})
+ \cos\theta \sin\varphi (e^{136} + e^{426})
+ \sin \varphi (e^{125} + e^{345}) 
\nonumber \\
& &  - \cos \varphi (e^{135} + e^{425})
- \sin \theta (e^{146} + e^{236})
\ ,
\label{jocp3} \\[2mm]
\im \Omega & = &
-\cos\theta \cos\varphi (e^{125} + e^{345})
- \cos\theta \sin\varphi (e^{135} + e^{425})
+ \sin \varphi (e^{126} + e^{346}) 
\nonumber \\
& & - \cos \varphi (e^{136} + e^{426})
+ \sin \theta (e^{145} + e^{235})
\ .
\nonumber
\eeqa 
These forms satisfy (\ref{su3}). 

The torsion classes are found after computing the exterior derivatives
that turn out to be exactly of the form (\ref{nonnk}) with
\beq
\cw_1 = \frac{2(1+\lambda^2)}{3\lambda}  \quad ; \quad 
\cw_2 \wedge J = \lambda J^2 - 6\lambda e^{1234}  \ .
\label{w1w2cp3} 
\eeq
Both $\cw_1$ and $\cw_2$ are real. In fact, $d\im \Omega=0$.
We can check that $\cw_2 \wedge J \wedge J = 0$ so that $\cw_2$ is primitive. 
It also follows that
\beq
{}^*(\cw_2 \wedge J) = 2\lambda J - 6\lambda e^{56} \ . 
\label{dualw2}
\eeq
With all this information it is a simple exercise to verify that the RR 2-form
$F_2$ given in (\ref{f2cp3}) can indeed be written as (\ref{f2nnk}) when $\lambda^2=\frac15$.

\section{Final remarks}
\label{conclu}

The original motivation behind this paper was to identify the internal space
implicit in a class of \neq1 type IIA $\ads_4$ vacua obtained using the effective
\deq4 formalism. As we have explained, this internal space turns out to be
$\S^3 \times \S^3/\Z_2^3$ with a nearly-K\"ahler metric. This property, together
with the structure of background fluxes, is in complete agreement with the
general results derived from supersymmetry conditions and equations of motion in
\deq10.  

Unlike the Minkowski case, $\ads_4$ \neq1 type IIA compactifications 
have the peculiarity that the equations of motion can be solved 
in the absence of orientifold planes of negative tension. 
In the \deq4 approach this can be simply understood from the tadpole cancellation equations
that include fluxes and sources \cite{cfi}. In \deq10, as reviewed in section 2,
this follows from the Bianchi identity for the RR 2-form \cite{gmpt2}. 
In the $\S^3 \times \S^3/\Z_2^3$ compactification we have found explicit
solutions of the tadpole cancellation conditions and used them to 
construct configurations of D6-branes that allow to solve the Bianchi identity
in \deq10.  

A second motivation of our work was to find a concrete example of 
\neq1 type IIA compactification to $\ads_4$ in which the internal space
is not nearly-K\"ahler. This possibility is allowed by the general analysis
of flux vacua, it corresponds to the case in which both torsion classes
$\cw_1$ and $\cw_2$ are different from zero. Previous attempts to 
construct examples of this sort failed because the Bianchi identity
for the RR 2-form could not be fulfilled \cite{lt}. Our contribution
has been to observe that some solutions of massless type IIA supergravity
discovered long time ago \cite{np, vst, stv, pp} do fit within the general framework
of $\ads_4$ flux vacua while the internal manifold does not have a nearly-K\"ahler
metric. We considered compactification on $\C\P^3$ and showed that both torsion
classes $\cw_1$ and $\cw_2$ are different from zero. Moreover, the background
of the RR 2-form has the correct expression in terms of the torsion classes.
Another example with both $\cw_1$ and $\cw_2$ non zero, already studied in \cite{hp},
which has as internal space the coset $SU(3)/U(1)^2$, can be treated along the same lines
as in section \ref{ccp3}.

In this note we have exemplified the validity and applicability
of the effective \deq4 approach to uncover properties of \deq10 flux
vacua. It is clearly desirable to extend the effective
formalism to compactifications with more generic internal manifolds.
In the future we hope to join efforts in pursuing further research
on this interesting subject.

\vspace*{1cm}

{\bf \large Acknowledgments}

We thank B. Acharya, M. Gra\~na, L. Ib\'a\~nez and J. Maldacena for
useful comments.
We are specially grateful to P. C\'amara and S. Theisen for clarifying
explanations and sharing their notes, and to Y. Oz for bringing \cite{np} to our attention.
A.F. acknowledges the hospitality of the Max-Planck-Institut f\"ur
Gravitationsphysik while preparing this paper.
This work has been partially supported by the European Commission under
the RTN European Program MRTN-CT-2004-503369, the
Comunidad de Madrid through Proyecto HEPHACOS S-0505/ESP-0346, and the 
AECI (Agencia Espa\~nola de Cooperaci\'on Internacional).
G.A. acknowledges the hospitality of the International Centre for
Theoretical Physics where part of this work was done.
The work of G.A. is partially supported by ANPCyT grant PICT 11064 and
CONICET PIP 5231.

\section*{Appendix A: Effective approach in \deq4}
\label{appA}
%\addcontentsline{toc}{section}{\hspace{13pt} Appendix A:  $SU(3)$ structure of twisted torus }
\setcounter{equation}{0}
\renewcommand{\theequation}{A.\arabic{equation}}

This appendix serves several purposes. First we give a concise account of the effective 
action for \deq4, \neq1 type IIA toroidal orientifolds \cite{gl,dkpz,vz}. 
We then describe to some extent the specific model that turns out to be related to
compactification on $\ads_4 \times \S^3 \times \S^3$. We will also show that the results  
are in complete agreement with those derived from supersymmetry conditions
and equations of motion in \deq10. Finally, we discuss tadpole cancellation equations and provide
configurations of supersymmetric D6-branes that solve these equations.

In the \deq4 effective formalism the analysis of vacua is based on the superpotential generated
by RR, NSNS and geometric fluxes. In type IIA orientifolds the flux induced superpotential 
can be written as
\beq
W = \int_{\M_6} \!\!\! e^{J_c} \wedge \ov{F} + \Omega_c \wedge (\ov{H} + dJ_c)  \ .
\label{fullw}
\eeq
The complexified forms defined as
\beq
J_c = B + iJ \quad ; \quad \Omega_c= C_3 + i\re (e^{-\phi} \Omega) \ ,
\label{cforms}
\eeq
are expanded in the basis of invariant 2 and 3-forms, with coefficients given
by the moduli fields. In the model we are considering these fields are the axiodilaton $S$,
together with three complex structure $U_i$ and three K\"ahler moduli $T_i$. 
The relevant expansions are
\beq
J_c = iT_i \omega_i  \quad ; \quad \Omega_c= iS \a_0 - iU_i \a_i  \ .
\label{cformexps}
\eeq
As we saw in section \ref{cs3s3}, $J=t_i \omega_i$. The NSNS 2-form can also be expanded 
in terms of the $\omega_i$ as $B=-v_i \omega_i$. The $v_i$ are the K\"ahler axions and
indeed the K\"ahler moduli are $T_i=t_i+i v_i$. For the axiodilaton and complex structure
moduli we can substitute (\ref{om6}) to obtain
\beq
\re S \equiv s = e^{-\phi} \, \sqrt{\frac{t_1t_2t_3}{\tau_1\tau_2\tau_3}} 
\quad ; \quad
\re U_i \equiv u_i = s \tau_j \tau_k \  , \   j \not= k \ .
\label{defrsu}
\eeq
The corresponding axions arise from the RR 3-form given by $C_3 = -\im S \a_0 + \im U_i \a_i$.

To compute the superpotential we need expansions for the background fluxes.
We follow the approach of \cite{cfi} where the fluxes are chosen to comply with the
$\Z_2 \times \Z_2$ symmetry (\ref{z2z2}). Thus, just as
$J_c$ and $\Omega_c$,  the fluxes are to be expanded in the basis
of invariant forms displayed in (\ref{allforms}). 
Furthermore, since we are assuming that the moduli are those of
toroidal IIA orientifolds, the fluxes are required to conform to the
orientifold involution (\ref{oaction}). This means that $\ov{F}_0$ and $\ov{F}_4$ are even,
whereas $\ov{H}$, $\ov{F}_2$ and $\ov{F}_6$ are odd under the orientifold involution \cite{gl}.
The upshot is that background fluxes have the following expansions   
\beq
\begin{array}{lclcl}
\ov{H}  = h_0 \b_0 + h_i \b_i  & \ ; \ & \ov{F}_0 = - M 
& \ ; \ &
\ov{F}_2 =  q_i \omega_i   \\
\ov{F}_4  = e_i \tilde \omega_i &  \ ; \ &
\ov{F}_6 = e_0 \a_0 \wedge \b_0 &  .
\end{array}
\label{fluxbg}
\eeq
The exterior derivative of these fluxes and the K\"ahler form $J$ are found using
(\ref{abis}) that define the internal space. 

The scalar potential of the moduli has the standard \neq1 expression
\beq
V= e^K \big\{\!\!\! \sum_{\Phi=S,T_i, U_i} \!\!\! (\Phi + \Phi^*)^2 |D_\Phi W|^2
- 3|W|^2 \big\} \ ,
\label{spot}
\eeq 
where we already assumed the classical K\"ahler potential 
$K = - \sum_{\Phi=S,T_i, U_i} \, \log(\Phi + \Phi^*)$,
and  $D_\Phi W = \partial_\Phi W + W \partial_\Phi K$. Supersymmetric AdS minima are determined by
the condition $D_\Phi W=0$.

To obtain the model analyzed in \cite{cfi} one chooses RR fluxes $q_i=-c$ and $e_i=e$ so that
a configuration with $T_i=T$ is allowed. The resulting superpotential is simply\footnote{A volume  factor ${\cc}$ appears here due to normalization (\ref{MGvolume}). }
\beq 
\frac{W}{\cc}=e_0 + 3ieT + 3c T^2 + iM T^3 +
(ih_0 - 3a T)S - \sum_{k=0}^3 (ih_k + b_kT) U_k\ .
\label{supT}
\eeq  

This superpotential admits supersymmetric AdS minima provided that the fluxes satisfy the constraint
\beq
3h_k a + h_0 b_k = 0
\quad ; \quad k=1,2,3 \ .
\label{homega}
\eeq
In this case the real parts of the axiodilaton and complex structure moduli are completely 
determined in terms of the K\"ahler modulus according to
\beq
as = 2t(c-Mv) \quad ; \quad 3 as = b_k u_k \quad ; \quad k=1,2,3 \ .
\label{realsu}
\eeq
Recall that $s=\re S$, $u_k=\re U_k$, $t=\re T$ and $v=\im T$ and that the real part of
the moduli are positive definite. Thus, (\ref{realsu}) requires that the geometric fluxes
$a$ and $b_k$ be of the same sign.    
For the $S$ and $U_i$ axions only one linear combination is fixed, this is
\beq
3a \im S +b_k \im U_k = 3e + \frac{3c}{a}(3h_0
- 7a v) - \frac{3M}{a} v(3h_0 - 8a v) \ .
\label{axionfix}
\eeq
To have the minimum with $T_i=T$ it must also be that
$b_1 \im U_1 = b_2 \im U_2 = b_3 \im U_3$.

The vacuum expectation value for the K\"ahler modulus depends on whether the mass parameter
$M$ vanishes or not.
When $M=0$ it is found that
\beq
v=v_0=\frac{h_0}{3a} \quad ; \quad 9c t^2 = e_0 - \frac{h_0 e}{a} -
\frac{h_0^2 c}{3a^2} \ .
\label{valmzero}
\eeq
In this case (\ref{realsu}) implies that
necessarily there is a flux of the RR 2-form, i.e. $c \not= 0$, and furthermore
that $ac > 0$ and $c b_k > 0$.
Background fluxes $\ov{H}$ and $\ov{F}_4$ can be absent but then the Freund-Rubin
flux $\ov{F}_6 \sim e_0$ must be turned on. 

When $M \not= 0$ the K\"ahler axion satisfies the cubic equation 
\beq 
160(v-v_0)^3 + 294(v_0-\frac{c}{\msm M})(v-v_0)^2 +  135(v_0-\frac{c}{\msm M})^2(v-v_0) + 
v_0^2(v_0-\frac{3c}{\msm M}) + \frac{1}{\msm{M}a}(e_0 a - e h_0) = 0 \ .
\label{cubiceq}
\eeq
The real part of the K\"ahler modulus is now determined from
\beq
t^2= 15 (v-\frac{c}{\msm M})(v-v_0) \ .
\label{valtm}
\eeq
 The solution for $v$ must be real and such that $t^2 > 0$.

Let us now check that the above results agree with the general analysis in \deq10.
To begin observe that we have $d \im \Omega=0$ compared to $d \re \widehat \Omega=0$. 
We find that in order to match the \deq4 and \deq10 results we need to make the choice 
\beq
\widehat \Omega = - i\Omega \quad ; \quad \a + \b=0 \, {\rm mod} \, 2\pi \ .
\label{wophase}
\eeq
The full exterior derivatives of $J$ and $\Omega$ are given in (\ref{djdo}). 

The next step is to express the field strengths in terms of the background fluxes and the
moduli.
In the case at hand, with $q_i=-c$, $e_i=e$, $T_i=T$, it is possible
to write most forms in terms of $J$ and $\Omega$. For example, $B=-v J/t$, $\ov{F}_4=e J^2/2t^2$,
and so on. After substituting in (\ref{hfdef}) we find
\beqa
H & = & \frac{s e^\phi}{t^3} (h_0 - 3 av) \im \Omega  \ , 
\nonumber \\
F_2 & = &\frac{Mv-c}{t} \, J   \ ,
\label{hfd4} \\
F_4 & = & \big[3e + 6cv -3Mv^2 - (3a \im S + b_i \im U_i) \big] \, \frac{J^2}{6t^2} \ ,
\nonumber \\
F_6 & = & \big[e_0 - 3ev - 3cv^2 + Mv^3 +(v-\frac{h_0}{3a}) (3a \im S + b_i \im U_i) \big] \,  
\frac{J^3}{6t^3} \ .
\nonumber 
\eeqa
All these expressions greatly simplify upon using (\ref{axionfix}) and (\ref{cubiceq}).
In the end we obtain
\beqa
\!\!\!\!\!\!\! dJ \!\!\!& = &\!\!\! \frac{6(c-Mv)}{t} \, e^{\phi} \im \Omega \quad ; \quad
d\Omega = \frac{4(c-Mv)}{t} \, e^{\phi} J^2  \quad ; \quad   H  =  \frac25 M \, e^\phi \im \Omega \ ,
\label{fd4} \\[2mm]
\!\!\!\!\!\! F_0 \!\!\!& = &\!\!\! -M  \quad ; \quad F_2 =   \frac{Mv-c}{t} \, J \quad ; \quad
F_4  =  -\frac{3M}{10} J^2 \quad ; \quad
F_6 =  \frac{3(c-Mv)}{2t} \, J^3 \ . 
\nonumber 
\eeqa
These agree with (\ref{derj}) and (\ref{fluxsol}) provided that 
\beq
\phi=3A \quad ; \quad m = \frac{M}5 \, e^{4A} \quad ; \quad  \tilde m = \frac{3(c-Mv)}{t} \, e^{4A} \ .
\label{checkft}
\eeq
The relation between the dilaton and the warp factor is precisely the same found in the ten-dimensional
analysis.

It is also interesting to compute the cosmological constant which
follows from the value $V_0$ of the scalar potential at the minimum. 
For the AdS minimum, $V_0 = -3 e^K |W_0|^2$. To determine $W_0$ we can substitute the vevs of the moduli 
in (\ref{supT}). A more general approach is to use the original form of the superpotential (\ref{fullw}).
Using previous results to evaluate the integrand at the minimum we arrive at
\beq
 e^{J_c} \wedge \ov{F} + \Omega_c \wedge (\ov{H} + dJ_c)  \big |_0 = 
\frac{2i}3 (m +i \tilde m) \, e^{-4A} \, J^3  \ .
\label{intcero}
\eeq 
This shows that the superpotential at the minimum is proportional
to the complex constant $\mu$. More precisely, $|W_0|^2=16 t^6 e^{-8A} |\mu|^2{\cc}^2$. 
For the classical K\"ahler potential, $e^K = (2^7 t^3 s u_1 u_2 u_3\,{\cc})^{-1}$, which
can be rewritten as $e^K=e^{4\phi}/128 t^9 {\cc}^3$.
Therefore, 
\beq
V_0 = -\frac{3e^{4A}}{8 {\cc}t^3} |\mu|^2 = \frac{\Lambda}{M_P^2} \ .
\label{ccd4}
\eeq 
where $\Lambda=-3 |\mu|^2$ is the cosmological constant and
$M_P^2 = 8 e^{2A-2\phi} {\cc}t^3$. The moduli above are evaluated at the minimum
and we are taking $\a^\prime=1$.

\subsection*{A.1 \ {\large{\bf Tadpole Cancellation and D6-branes}}}
\label{tadin4}

The fluxes induce tadpoles for the RR 7-form $C_7$ that can also couple to
D6-branes and O6-planes. In general these objects span space-time and wrap
a 3-cycle in $\M_6$. 
The RR 7-form can then be written as 
$C_7={\rm dvol}_4 \wedge X$, where $X$ is some 3-form in
the internal space, which can be expanded in a convenient basis.
We denote by $\Pi_a$ the 3-cycle wrapped by a stack of
$N_a$ $\D6_a$-branes or $\O6_a$-planes. The coupling of $C_7$ in the action has
contributions from fluxes and from the sources, namely
\beq
\int_{\M_4 \times \cs_6}\hspace{-5mm} C_7 \wedge (d\ov{F}_2 - F_0 \ov{H}) +
\sqrt{\cc}\, \sum_a N_a Q_a \int_{\M_4 \times \Pi_a} \hspace{-5mm} C_7 \ ,
\label{c7tadpole}
\eeq
where $Q_a=1$ for D6-branes and  $Q_a=-4$ for O6-planes.
Here we are considering the internal space to be $\cs_6=\S^3\times \S^3/\Z_2^3$.
The factor $\sqrt{\cc}$ must be included  for consistency with the normalization 
of the 1-form basis (see \ref{MGvolume}).

As usual in the \deq4 effective formulation, it appears useful to  consider factorizable 
3-chains
\beq
\Pi_a=(n_a^1, m_a^1)\otimes(n_a^2, m_a^2) \otimes(n_a^3, m_a^3) \  ,
\label{facpi}
\eeq
where $n_a^i$ $(m_a^i)$ are the wrapping numbers along the $\eta^i$ $(\eta^{i+3})$. 
In particular, there is a basis of 3-chains $\Pi_{ijk}$ spanning the $\{i,j,k\}$
directions. 
For instance, $\Pi_{123}=(1,0)\otimes(1,0) \otimes(1,0)$, etc.. To each $\Pi_{ijk}$  
we have a corresponding ``dual'' 3-form  $\eta^{ijk}$ such that 
\begin{equation}
 \int_{ \Pi_{i^{\prime}j^{\prime}k^{\prime}}}\eta^{ijk}= \frac{1}{\sqrt{\cc}}\, \delta_{i,i^{\prime}}
\, \delta_{j,j^{\prime}}\, \delta_{k,k^{\prime}} \ .
\label{intetaijk}
\end{equation}
Integrating over the 3-chain $\Pi_a$ then gives,
$\int_{ \Pi_a}\eta^{123}=\frac{1}{\sqrt{\cc}}  \, n_a^1 n_a^2 n_a^3$,\, 
$\int_{ \Pi_a}\eta^{156}=\frac{1}{\sqrt{\cc}}  \, n_a^1 m_a^2 m_a^3$, and so on.

It is worth noticing that the basis manifolds $\Pi_{ijk}$  are not necessarily closed cycles and, 
therefore, neither is $\Pi_a$, for generic wrappings. As an example, consider the exact form
$d (\xi^1 \wedge {\hat  \xi}^1) = 2 \sqrt{ab_1b_2b_3}  
(a \eta^{456}+ b_1\eta^{423}-b_2 \eta^{153}-b_3 \eta^{126})$, then,   
$ \int_ {\Pi_{456}}d (\xi^1 \wedge \xi^1)=\frac{2}{\sqrt{\cc}}  \,\sqrt{ab_1b_2b_3}$, 
indicating that the manifold $\Pi_{456}$ has a boundary (see \cite{marche} for a related discussion).
We rely on tadpole cancellation and supersymmetry to restrict to the adequate wrappings
for D6-branes. 
When the orientifold action (\ref{oaction}) is implemented there are eight 
O6-planes along $\otimes_i(1,0)$ and image D6-branes wrapping  $\otimes_i (n_a^i, -m_a^i)$ 
must be included.

To preserve the same supersymmetries as the background the D6-branes must
wrap cycles $\Pi_a$ such that
\begin{equation}
\theta_a^1+\theta_a^2+\theta_a^3=0\quad{\rm mod}\quad 2 \pi  \ ,
\label{susyd6}
\end{equation} 
where the angles are measured in accordance with
\beq
\tan \theta^j_a = \frac { m^j_a \tau_j}{ n^j_a } \ .
\label{tans}
\eeq
Recall that the $\tau_i$ are the complex structure moduli that enter in the metric
as shown in (\ref{metricm6}). In the vacuum we are considering they satisfy (\ref{crux}).

>From the supersymmetric constraint (\ref{susyd6}) it follows that
\begin{equation}
\tau_1\tau_2\tau_3 m_a^1 m_a^2 m _a^3-\tau_1 m_a^1 n_a^2 n_a^3- \tau_2 n_a^1
m_a^2 n_a^3 - \tau_3 n_a^1n_a^2 m_a^3 = 0  \ .
\label{tg}
\end{equation} 
This condition amounts to $\im \Omega \big |_{\Pi_a} =0$.
In fact, the supersymmetric cycles are calibrated by $\re \Omega$
and the condition on the angles is equivalent to  $\re \Omega \big |_{\Pi_a} = {\rm dvol}(\Pi_a)$.
Besides, the factorizable cycles satisfy $J \big |_{\Pi_a} =0$.

Substituting the fluxes and the data for the sources in (\ref{c7tadpole}) we obtain
the tadpole cancellation equations. The conditions receiving flux contributions are 
\beqa
\sum_a N_a Q_a n_a^1 n_a^2 n_a^3 +(M h_0 -3a c)  & = &  0 \ , \nonumber \\[0.2cm]
\sum_a N_a Q_a n_a^1 m_a^2 m_a^3 + (M h_1 + b_1 c) & = & 0 \ , \label{tadb}  \\[0.2cm]
\sum_a N_a Q_a m_a^1 n_a^2 m_a^3 + (M h_2 + b_2 c) & = & 0 \ , \nonumber \\[0.2cm]
\sum_a N_a Q_a m_a^1 m_a^2 n_a^3 + (M h_3 + b_3 c) & = & 0 \ . \nonumber
\eeqa
The sum in $a$ includes $\O6_a$-planes, when orientifold actions
are performed, as well as \mbox{$\D6_a$-branes} and their orientifold images if necessary.
Finally, there are fluxless conditions
\beqa
\sum_a N_a Q_a m_a^1 m_a^2 m_a^3 & = & 0 \ , \nonumber \\[0.2cm]
\sum_a N_a Q_a m_a^1 n_a^2 n_a^3 & = & 0 \ , \label{tadfless} \\[0.2cm] 
\sum_a N_a Q_a n_a^1 m_a^2 n_a^3 & = & 0 \ ,  \nonumber \\[0.2cm]
\sum_a N_a Q_a n_a^1 n_a^2 m_a^3 & = & 0 \ . \nonumber
\eeqa  
When the orientifold action (\ref{oaction}) is implemented these last four
constraints are automatically satisfied once images are included.

When $M \not=0$ the tadpole equations admit a solution without branes or O-planes. 
This happens because $h_k=-h_0 b_k /3a$ and then all flux tadpoles can cancel simultaneously
when $Mh_0 = 3ac$ \cite{cfi}. Now, the relations (\ref{checkft}) and (\ref{valtm})
imply that this condition is equivalent to $\tilde m^2 = 15 m^2$.
As explained in section \ref{d10} this is precisely the case when the internal space is 
nearly-K\"ahler and no sources are necessary to satisfy the Bianchi identity for $F_2$.
In \deq10 we have further seen that when $\tilde m^2 > 15 m^2$ the Bianchi identity
can be satisfied by adding sources of positive charge.
In the \deq4 approach it is indeed evident that whenever $Mh_0 < 3ac$ the tadpoles can
be cancelled by adding only $\D6$-branes. 

\begin{table}[htb] \footnotesize
\renewcommand{\arraystretch}{1.25}
\begin{center}
\begin{tabular}{|c||c|c|c|}
\hline $N_a$ & $(n_a^1,m_a^1)$ & $(n_a^2,m_a^2)$ & $(n_a^3,m_a^3)$ \\
 \hline\hline $N_A$ & $(1,0)$ & $(1,0)$ & $(1,0 )$ 
 \\ $N_B$ & $(-1,1)$ & $ (-1,1)$ & $(-1,1)$ \\
 $N_B$ &  $(-1,-1)$ & $ (-1,-1)$ & $(-1,-1)$ 
 \\ \hline 
\end{tabular}
\end{center} 
\caption{\small 
Wrapping numbers for D6-branes in model 1}
\label{adsm0}
\end{table}
 
To present examples of tadpole cancellation with only D6-branes we will consider 
the case $M=0$ in the $\S^3 \times \S^3$ compactification that we have been analyzing. 
A first model consists of the factorizable D6-branes shown in table \ref{adsm0}. 
The third stack is the mirror image, $\tilde m_B^i= -m_B^i$, of the second
and it is included to saturate the fluxless tadpole equations. We also take $N_A=8 N_B$.
The first stack has $\theta_A^i=0$, hence it 
is supersymmetric independently of the values of the complex structure parameters.   
On the other hand, substituting the wrapping numbers in (\ref{tadb}) gives the relations
$2N_B=ac=b_1c=b_2c=b_3c$, needed for tadpole cancellation.
Next, using that $\tau_i = b_i \sqrt{3 a/b_1 b_2 b_3}$, we find $\tau_1=\tau_2=\tau_3=\sqrt3$.
We can then check that the second and third stack are also supersymmetric.
In fact, computing $\tan \theta_B^i$ for the second shows that
$\theta_B^i=2\pi/3$. 
Assuming that the $\Pi_a$ cycles have large volume, 
in this model 1 the resulting gauge group is $U(N_A) \times U(N_B) \times U(N_B)$.
According to the intersections between cycles, the matter content consists of chiral multiplets 
transforming as
\beq
({\bm{N_A}},{\bm{\ov{N}_B}},{\bf 1}) + ({\bm{\ov{N}_A}},{\bf 1},{\bm{N_B}}) 
+ 8({\bf 1},{\bm{N_B}},{\bm{\ov{N}_B}}) \ .
\label{specuno}
\eeq 
The multiplicity 8 of the last representation arises from the intersection number between the
cycle B and its mirror. Since $N_A=8N_B$ this chiral spectrum is free of gauge anomalies.

\begin{table}[htb] \footnotesize
\renewcommand{\arraystretch}{1.25}
\begin{center}
\begin{tabular}{|c||c|c|c|}
\hline $N_a$ & $(n_a^1,m_a^1)$ & $(n_a^2,m_a^2)$ & $(n_a^3,m_a^3)$ \\
 \hline\hline $N_0$ & $(1,0)$ & $(1,0)$ & $(1,0 )$ 
 \\ $N_1$ & $(1,0)$ & $ (0,-1)$ & $(0,1)$ \\
 $N_2$ &  $(0,1)$ & $ (1,0)$ & $(0,-1)$ \\
 $N_3$ &  $(0,1)$ & $ (0,-1)$ & $(1,0)$
 \\ \hline 
\end{tabular}
\end{center} 
\caption{\small 
Wrapping numbers for D6-branes in model 2}
\label{oadsm0}
\end{table}

There are other D6-brane configurations capable of canceling tadpoles. 
Some are equivalent to the setup in model 1 but others belong to a different class. 
For instance, in table \ref{oadsm0}
we display a model 2 with four stacks of branes that are all supersymmetric independently of
the complex structure moduli. To cancel tadpoles the numbers of branes in each stack
must be related to the fluxes by $N_0=3ac$ and $N_i = b_ic$. In this model the resulting
spectrum is non-chiral.

\section*{Appendix B: Supersymmetric vacua of 
massless type IIA supergravity in \deq10}
\label{appB}
%\addcontentsline{toc}{section}{\hspace{13pt} Appendix B: }
\setcounter{equation}{0}
\renewcommand{\theequation}{B.\arabic{equation}}
 
In this appendix we tersely summarize some basic aspects of compactification of
massless IIA supergravity on ${\rm AdS}_4 \times \M_6$. We will review the case
when the internal space is $\C\P^3$ and appropriate fluxes are turned on so that
there is a vacuum with \neq1 supersymmetry in \deq4 \cite{np,vst,stv}. Our main goal
is to explicitly find the six dimensional Killing spinor in order to determine
the fundamental forms $J$ and $\Omega$ that define the $SU(3)$ structure. 
We will follow and refer to the discussion of \cite{np} where the equations of
motion and the supersymmetry transformations are spelled out in full detail.

We consider a class of vacua with background metric of the product form (\ref{geo}) but to 
simplify the warp factor $A$ is fixed to zero. The dilaton $\phi$ is assumed to be constant whereas    
the NS 2-form and its field strength are taken to vanish. On the contrary, there are
non-trivial RR fluxes. For the 4-form one makes the Freund-Rubin Ansatz
\beq
F_{\mu \nu \a \b} = 3 f \epsilon_{\mu \nu \a \b} \quad ; \quad 
\epsilon_{0123}=\sqrt{-g_4}   \ ,
\label{marc}
\eeq
while other components are zero. For the RR 2-form there is a flux $F_{mn}$ through $\M_6$ to be specified
shortly. Under these conditions the equations of motion reduce to
\beqa
R_{\mu \nu} & = & -12 e^{\phi/2} f^2 g_{\mu \nu}  \ , \nonumber \\[2mm]
R_{mn} & = & 6 e^{\phi/2} f^2 g_{m n} + \oh e^{3\phi/2} F_{mp} F_n^{\  p}  \ , \label{eommz} \\[2mm]
e^\phi  F_{mn} F^{mn} & = & 24 f^2  \quad ; \quad   \nabla_m F^{mn}=0  \  .
\nonumber
\eeqa
The Einstein equation in \deq4 shows that space-time can indeed be taken to be ${\rm AdS}_4$
with cosmological constant $\Lambda= -12 e^{\phi/2} f^2 $.

To characterize the internal space we still need to specify the flux $F_2$. We will see that it
is consistent to take $\M_6$ to be $\C\P^3$ with metric given in (\ref{cp3metric}), while $F_2$ 
can be set equal to the 2-form (\ref{f2cp3}). This RR 2-form satisfies the equation of motion
and the properties
\beq
F_{mn} F^{mn} = 2(2\lambda^2 + \frac1{\lambda^2})   \quad ; \quad   F_{ac} F_b^{\  c}= \lambda^2 \d_{ab}
\quad ; \quad   F_{ik} F_j^{\  k}= \frac1{\lambda^2} \d_{ij} \ ,
\label{f2props}
\eeq
where $a,b,c=1, \cdots, 4$, and $i,j,k=5,6$, are flat indices.

Once the flux $F_2$ is given, the dilaton equation of motion implies that the vevs $e^\phi$, $f$ and
the metric parameter $\lambda$ are related by
\beq
e^\phi(2\lambda^2 + \frac1{\lambda^2}) = 12 f^2  \ .
\label{flambda}
\eeq    
Substituting in the \deq6 Einstein equation we then find that
in the flat basis the Ricci tensor is diagonal with components
\beq
R_{ab}=\oh e^{3\phi/2} (3\lambda^2+\frac1{\lambda^2}) \, \d_{ab}  \quad ; \quad 
R_{ij}= e^{3\phi/2} (\lambda^2 + \frac1{\lambda^2}) \, \d_{ij} \  .
\label{riccid6}
\eeq
The Ricci tensor of the generic $\C\P^3$ metric has precisely this structure. Comparing with
(\ref{riccicp3}) we see that the dilaton vev has to be $e^\phi=1$. Moreover, the parameter $\lambda$
must be such that 
\beq
5 \lambda^2 - 6 + \frac1{\lambda^2} = 0  \ .
\label{lequ}
\eeq
There is a solution with $\lambda^2=1$ for which the metric is Einstein. We are more interested
in the solution with $\lambda^2=1/5$. In this case, from (\ref{flambda}) it transpires that the Freund-Rubin 
parameter is fixed to be $f^2=9/20$. 

We now discuss the requirements for residual supersymmetry in \deq4. We will employ exactly  
the same conventions of \cite{np} for the \deq10 Dirac matrices. In \deq6 we basically adopt the matrices
used in \cite{dnp} in \deq7. With flat indices these are
\beqa
\Gamma_1 & = &  i \gamma_0 \otimes \uno  \quad ; \quad \Gamma_2  =  \gamma_1 \otimes \uno 
\quad ; \quad  \Gamma_3  =  \gamma_2 \otimes  \uno  \quad ; \quad \Gamma_4  =  \gamma_3 \otimes \uno 
\\[2mm]
\Gamma_5  & = &  i\gamma_5 \otimes \sigma_1 
\quad ; \quad  \Gamma_6  =  i\gamma_5 \otimes \sigma_2    \quad ; \quad 
\Gamma_0 = \Gamma_ 1 \cdots \Gamma_6 =   i\gamma_5 \otimes \sigma_3  \ ,
\nonumber
\label{gammad6}
\eeqa
 where $\sigma_i$ are the Pauli matrices. The 4-dimensional matrices are
\beq
\gamma_0 = 
\renewcommand{\arraystretch}{0.8}
\inpar{\!\!\begin{array}{cc}
0 & 1 \\
1 & 0 
\end{array}\!\!}
\quad ; \quad  
\gamma_a = 
\renewcommand{\arraystretch}{0.8}
\inpar{\!\!\begin{array}{cc}
0 & \sigma_a \\
-\sigma_a & 0 
\end{array}\!\!}  \ , 
\label{gammad4}
\eeq
and $\gamma_5= - i \gamma_0 \gamma_1 \gamma_2 \gamma_3$. We will also need the charge conjugation
matrix in \deq6 which in our conventions is given by $C=\Gamma_2 \Gamma_  4\Gamma_6$.

To study the conditions for the vacuum to preserve supersymmetry we first write the 10-dimensional parameter
as $\epsilon \otimes \eta$, where $\epsilon$ and $\eta$ are respectively spinors in four and six dimensions. 
We then substitute the vevs of all fields in the supersymmetry transformations of the fermionic fields which
in \deq10, IIA supergravity are the gravitino and the dilatino. We refer the reader to reference \cite{np}
for the explicit equations of these transformations. 
{}From the dilatino variation we obtain 
\beq
(S + 2f)\eta = 0  \ ,
\label{etaev}
\eeq
where the matrix $S$ depends on the RR 2-form flux as
\beq
S  = \oh F_{mn} \Gamma^{mn} \Gamma_0  \ .
\label{sdef}
\eeq
For the $F_2$ background in (\ref{f2cp3}), $S$ turns out to have eigenvalues
$\msm{1/\lambda}$, $\msm{(2\lambda^2-1)/\lambda}$, and  $-\msm{(2\lambda^2+1)/\lambda}$,
with degeneracies $4,2$ and 2 respectively. Remarkably, for the case of interest with
$\lambda^2=1/5$ and $f^2=9\lambda^2/4$, $S$ can have an eigenvalue $-2f$ as long as we
take $f=3\lambda/2$. The corresponding eigenvector has the simple form
\beq
\eta =
\renewcommand{\arraystretch}{0.8}
\inpar{\!\!\begin{array}{c}
s_1\\
s_2 \\
s_3 \\
s_4 \\
0\\
0\\
0\\
0 
\end{array}\!\!}
\quad ; \quad s_1 = -\frac{\sin \theta \, e^{-i\phi}\, s_3}{1 + \cos \theta}  
\quad ; \quad s_4 = \frac{\sin \theta \, e^{i\phi}\, s_2}{1 + \cos \theta} \ ,
\label{etafull}
\eeq
where $s_2$ and $s_3$ in principle depend on all internal coordinates.

{}From the gravitino variation $\delta \psi_\mu$, using (\ref{etaev}), we find
\beq
D_\mu \epsilon - e^{\phi/4} f \gamma_5 \gamma_\mu \epsilon = 0 \ .
\label{killingads}
\eeq
This is the expected equation for the supersymmetry parameter in 
${\rm AdS}_4$ with cosmological constant $\Lambda= -12 e^{\phi/2} f^2 $.
Finally, from the variation $\delta \psi_m$ we obtain the Killing equation
\beq
D_m \eta - \frac{f}2 \Gamma_m \eta - \frac14 F_m^{\ n} \Gamma_n \Gamma_0 \eta = 0 \ ,
\label{killingcp3}
\eeq
where we have set  $e^\phi=1$. For the covariant derivative acting on spinors
we use the conventions of \cite{dnp}.

It remains to solve the Killing equation to determine the unknown functions $s_2$ and
$s_3$ in $\eta$. From the $\psi$ component we find
\beq
s_2 = i  e^{-i\phi}\, s_3  \ .  
\label{s2s3}
\eeq
It further follows that $s_3$ is completely independent of the $\S^4$ coordinates
$(\psi, \alpha, \beta. \gamma)$, but depends on the $\S^2$ variables as
\beq
s_3 = e^{i\delta}  \, e^{-i\phi/2} \, \cos \frac{\theta}2  \ ,
\label{s3final}
\eeq
where $\delta$ is a constant phase. The normalization guarantees that the Weyl spinors 
\beq
\eta_\pm = \frac{1 \pm i\Gamma_0}2 \, \eta 
\label{etaweyl}
\eeq
satisfy $\eta_\pm^{\dagger} \eta_\pm =1$. The phase $\d$ is fixed by imposing the reality condition 
$\eta_+^* = C \, \eta_-$.

We are now ready to compute the fundamental forms $J$ and $\Omega$ defined by
\beq
J_{mn} = i \eta_-^\dagger \, \Gamma_{mn} \, \eta_-   \quad ; \quad     
\Omega_{mnp} = \eta_+^\dagger \, \Gamma_{mnp} \, \eta_-  \ .
\label{joeta}
\eeq
In the end we obtain the results reported in section \ref{ccp3}. We stress that there is a unique
Killing spinor $\eta$ so that the internal manifold has $SU(3)$ structure and there is \neq1
supersymmetry in \deq4.

\end{document}